\documentstyle[12pt]{article}
\begin{document}
\pagestyle{plain}
\setcounter{page}{1}
\baselineskip16pt

\begin{titlepage}

\begin{flushright}
PUPT-1521\\
hep-ph/9411430
\end{flushright}

\vspace{20 mm}

\begin{center}
{\huge Kaon Condensation in the Bound-State}\par
\vspace{5mm}
{\huge Approach to the Skyrme Model}
\end{center}

\vspace{10 mm}

\begin{center}
{\large Karl M.\ Westerberg}\par
\vspace{3mm}
Joseph Henry Laboratories\\
Princeton University\\
Princeton, New Jersey 08544
\end{center}

\vspace{2cm}

\begin{center}
{\large Abstract}
\end{center}

We explore kaon condensation using the bound-state approach to the
Skyrme model on a 3-sphere.  The condensation occurs when the energy
required to produce a $K^-$ falls below the electron fermi level.
This happens at the baryon number density on the order of 3--4 times
nuclear density.

\vspace{2cm}

\begin{flushleft}
November 27, 1994
\end{flushleft}

\end{titlepage}
\newpage

\renewcommand{\baselinestretch}{1.1}  


%
%
%
%
%
%
%
%
%
%
%


\newcommand{\AAA}[3]{A_{#1} A_{#2} A_{#3}}
\newcommand{\commAA}[2]{\left[A_{#1},A_{#2}\right]}
\newcommand{\DK}[1]{D_{#1}K}
\newcommand{\dK}[1]{\partial_{#1}K}
\newcommand{\hc}[1]{{#1}^\dagger}
\newcommand{\leftparens}[1]{\left( #1 \right.}
\newcommand{\mAA}[2]{A_{#1} A_{#2}}
\newcommand{\mtxIIIxIII}[4]{\parens{
   \begin{array}{c|c}
      {\displaystyle \vphantom{\underline{#1}} #1}
         & {\displaystyle \vphantom{\underline{#2}} #2} \\ \hline
      {\displaystyle \vphantom{\sqrt{#3}} #3}
         & {\displaystyle \vphantom{\sqrt{#4}} #4}
   \end{array}
}}
\newcommand{\noparens}[1]{#1}
\newcommand{\parens}[1]{\left( #1 \right)}
\newcommand{\power}[2]{{#1}^{#2}}
\newcommand{\rightparens}[1]{\left. #1 \right)}
\newcommand{\taudot}[1]{#1 \cdot \vec{\tau}}
\newcommand{\trAAA}[3]{tr\parens{\AAA{#1}{#2}{#3}}}
\newcommand{\trAA}[2]{tr\parens{\mAA{#1}{#2}}}
\newcommand{\Flagprefactor}{\frac{\pi f_\pi}{e}\int_{-\infty}^{\infty}dx\,}
\newcommand{\klagprefactor}{\hc{\chi}\chi
   \frac{\pi}{ef_\pi} \int_{-\infty}^{\infty}dx\,}

\newcommand{\leftparensI}[1]{( #1}
\newcommand{\leftparensII}[1]{\Big( #1}
\newcommand{\leftparensIII}[1]{\bigg( #1}
\newcommand{\rightparensI}[1]{#1 )}
\newcommand{\rightparensII}[1]{#1 \Big)}
\newcommand{\rightparensIII}[1]{#1 \bigg)}
\newcommand{\parensI}[1]{( #1 )}
\newcommand{\parensII}[1]{\Big( #1 \Big)}
\newcommand{\parensIII}[1]{\bigg( #1 \bigg)}

\def\pnum#1{}

\newcommand{\FlagtermI}{\pnum{I}
\alpha
}

\newcommand{\FlagtermII}{\pnum{II}
\leftparensII{\frac{\sin^{2}{F}}{1+\power{x}{2}}}
}

\newcommand{\FlagtermIII}{\pnum{III}
+{\rightparensII{\frac{1}{2} \power{\parens{F'\parens{x}}}{2}}}
}

\newcommand{\FlagtermIIII}{\pnum{IIII}
+{\frac{1}{\alpha}}
}

\newcommand{\FlagtermV}{\pnum{V}
\leftparensII{2 \sin^{4}{F}}
}

\newcommand{\FlagtermVI}{\pnum{VI}
+{\rightparensII{4 \parensI{1+\power{x}{2}} \sin^{2}{F}
\power{\parens{F'\parens{x}}}{2}}}
}

\newcommand{\klagtermI}{\pnum{I}
\power{\parens{k'\parens{x}}}{2}
}

\newcommand{\klagtermII}{\pnum{II}
\leftparensIII{4 \alpha}
}

\newcommand{\klagtermIII}{\pnum{III}
+{\rightparensIII{\frac{8}{\alpha} \parensI{1+\power{x}{2}} \sin^{2}{F}}}
}

\newcommand{\klagtermIIII}{\pnum{IIII}
+{\frac{24}{\alpha} \parensI{1+\power{x}{2}}
\parens{1+\cos{F}} \sin{F}\, F'\parens{x} k\parens{x} k'\parens{x}}
}

\newcommand{\klagtermV}{\pnum{V}
-{\frac{4 \power{\omega}{2}}{\power{e}{2} \power{f_\pi}{2}}
\power{\parens{k\parens{x}}}{2}}
}

\newcommand{\klagtermVI}{\pnum{VI}
\leftparensIII{\frac{\power{\alpha}{3}}{\power{\parens{1+\power{x}{2}}}{2}}}
}

\newcommand{\klagtermVII}{\pnum{VII}
+{\alpha}
}

\newcommand{\klagtermVIII}{\pnum{VIII}
\leftparensII{\power{\parens{F'\parens{x}}}{2}}
}

\newcommand{\klagtermVIIII}{\pnum{VIIII}
+{\rightparensIII{\rightparensII{2 \frac{\sin^{2}{F}}{1+\power{x}{2}}}}}
}

\newcommand{\klagtermX}{\pnum{X}
+{\frac{4 e \omega N}{f_\pi \power{\pi}{2}}
\sin^{2}{F}\, F'\parens{x} \power{\parens{k\parens{x}}}{2}}
}

\newcommand{\klagtermXI}{\pnum{XI}
+{\power{\parens{k\parens{x}}}{2}}
}

\newcommand{\klagtermXII}{\pnum{XII}
\leftparensIII{\alpha}
}

\newcommand{\klagtermXIII}{\pnum{XIII}
\leftparensII{-{\power{\parens{F'\parens{x}}}{2}}}
}

\newcommand{\klagtermXIIII}{\pnum{XIIII}
+{2 \frac{\power{\parens{1+\cos{F}}}{2}}{1+\power{x}{2}}}
}

\newcommand{\klagtermXV}{\pnum{XV}
-{\rightparensII{2 \frac{\sin^{2}{F}}{1+\power{x}{2}}}}
}

\newcommand{\klagtermXVI}{\pnum{XVI}
+{\frac{8}{\alpha}}
}

\newcommand{\klagtermXVII}{\pnum{XVII}
\leftparensII{\frac{1}{4} \parensI{1+\power{x}{2}}
\power{\parens{1+\cos{F}}}{2} \power{\parens{F'\parens{x}}}{2}}
}

\newcommand{\klagtermXVIII}{\pnum{XVIII}
-{2 \parensI{1+\power{x}{2}} \sin^{2}{F}\, \power{\parens{F'\parens{x}}}{2}}
}

\newcommand{\klagtermXVIIII}{\pnum{XVIIII}
+{\power{\parens{1+\cos{F}}}{2} \sin^{2}{F}}
}

\newcommand{\klagtermXX}{\pnum{XX}
-{\rightparensII{\sin^{4}{F}}}
}

\newcommand{\klagtermXXI}{\pnum{XXI}
+{\rightparensIII{\frac{4 m_K^2 \power{\alpha}{3}}{\power{e}{2}
\power{f_\pi}{2}} \frac{1}{\power{\parens{1+\power{x}{2}}}{2}}}}
}

\newcommand{\FeqntermI}{\pnum{I}
\frac{1}{2} F''\parens{x}
}

\newcommand{\FeqntermII}{\pnum{II}
-{\frac{\cos{F} \sin{F}}{1+\power{x}{2}}}
}

\newcommand{\FeqntermIII}{\pnum{III}
+{\frac{4}{\power{\alpha}{2}}}
}

\newcommand{\FeqntermIIII}{\pnum{IIII}
\leftparensII{\parensI{1+\power{x}{2}} \sin^{2}{F}\, F''\parens{x}}
}

\newcommand{\FeqntermV}{\pnum{V}
+{\parensI{1+\power{x}{2}} \cos{F} \sin{F}\, \power{\parens{F'\parens{x}}}{2}}
}

\newcommand{\FeqntermVI}{\pnum{VI}
+{2 x \sin^{2}{F}\, F'\parens{x}}
}

\newcommand{\FeqntermVII}{\pnum{VII}
-{\rightparensII{\cos{F} \sin^{3}{F}}}
}


\newcommand{\refeqn}[1]{(\ref{#1})}

\renewcommand{\epsilon}{\varepsilon}
\newcommand{\SU}[1]{\mbox{${\rm SU}(#1)$}}
\newcommand{\WZ}{{\rm WZ}}
\newcommand{\QCD}{{\rm QCD}}
\newcommand{\disk}{{\rm D}}
\newcommand{\grad}{\nabla}
\newcommand{\MeV}{{\rm MeV}}
\newcommand{\fm}{{\rm fm}}
\newcommand{\rightvbar}[1]{\left. #1 \right|}
\newcommand{\tr}{\mathop{\rm tr}}
\newcommand{\sym}{\mathop{\rm sym}}
\newcommand{\arccot}{\mathop{\rm arccot}}
\newcommand{\order}{\mathop{\rm order}}
\newcommand{\id}{1}
\newcommand{\mtxIIxI}[2]{\parens{
   \begin{array}{c}
      #1 \\
      #2
   \end{array}
}}
\newcommand{\mtxIIxII}[4]{\parens{
   \begin{array}{cc}
      #1 & #2 \\
      #3 & #4
   \end{array}
}}

\newcommand{\sA}{{\cal A}}
\newcommand{\sB}{{\cal B}}
\newcommand{\sD}{{\cal D}}
\newcommand{\sL}{{\cal L}}
\newcommand{\sM}{{\cal M}}

\newcommand{\va}{\vec{a}}
\newcommand{\vb}{\vec{b}}
\newcommand{\vn}{\vec{n}}
\newcommand{\vv}{\vec{v}}
\newcommand{\vw}{\vec{w}}
\newcommand{\vx}{\vec{x}}
\newcommand{\vJ}{\vec{J}}
\newcommand{\valpha}{\vec{\alpha}}
\newcommand{\dalpha}{\dot{\valpha}}
\newcommand{\vtau}{\vec{\tau}}

\newcommand{\bc}{\bar{c}}
\newcommand{\bs}{\bar{s}}
\newcommand{\bnu}{\bar{\nu}}
\newcommand{\bomega}{\bar{\omega}}

\newcommand{\unitc}{\widehat{c}}
\newcommand{\unitr}{\widehat{r}}
\newcommand{\unitrho}{\widehat{\rho}}
\newcommand{\unittheta}{\widehat{\theta}}
\newcommand{\unitphi}{\widehat{\phi}}

\newcommand{\dc}{{\rm\bf d}c}
\newcommand{\dr}{{\rm\bf d}r}
\newcommand{\drho}{{\rm\bf d}\rho}
\newcommand{\dtheta}{{\rm\bf d}\theta}
\newcommand{\dphi}{{\rm\bf d}\phi}
\newcommand{\dt}{{\rm\bf d}t}

\newcommand{\fpi}{f_\pi}

\newcommand{\specialset}[1]{
   \hbox to 0pt{\sf #1\hss}
   \kern+.8pt
   \hbox{\sf #1}
}
\newcommand{\naturals}{\specialset N}
\newcommand{\integers}{\specialset Z}
\newcommand{\rationals}{\specialset Q}
\newcommand{\posreals}{\specialset P}
\newcommand{\reals}{\specialset R}
\newcommand{\complexes}{\specialset C}

\section{Introduction} \label{sect:1}

Recently it was discovered \cite{ref:15} that kaon condensation
in nuclear matter of densities on the order of a few times that of
normal nuclear matter may have an impact on the formation of low mass
black holes instead of neutron stars for masses on the order of 1.5
solar masses.

Since the proposal of kaon condensation and its possible consequences
in 1986 \cite{ref:16}, various attempts have been made to determine
the critical density at which kaon condensation sets in, by means of
a chiral lagrangian \cite{ref:16,ref:2,ref:1,ref:13} and also by
means of phenomenological off-shell meson-nucleon interactions
\cite{ref:3}.

In this paper, we explore kaon condensation within the framework of
the bound-state approach to the Skyrme model \cite{ref:6,ref:7}.  In
1985, Callan and Klebanov proposed the bound-state approach to the
Skyrme model \cite{ref:6} as a model for exploring the properties of
strange baryons.  To this day, the bound-state approach to the Skyrme
model remains one of the most successful models for hyperons.  At
around this time, it was suggested that the \SU{2} Skyrme model may
be used to determine the properties of dense nuclear matter by means
of putting skyrmions on a cubic lattice \cite{ref:8}.  It was later
determined that these \SU{2} skyrmions undergo a transition to a
smoothed out ``dissolved'' phase with restored chiral symmetry.
\cite{ref:14,ref:10}.  The cubic lattice symmetry was also replaced
by other symmetries which resulted in a lower crystal interaction
energy without affecting the qualitative phase structure of the
crystal \cite{ref:11}.\footnote{Unfortunately, kinetic energy effects
have not been computed with the Skyrme model due to its
non-renomalizability, so such comparison of energies is inherently
difficult.}

Unfortunately, lattice calculations can be difficult and require
involved numerical work.  In 1987, Manton proposed an alternative
to the lattice.  He put a single skyrmion on the surface of a
3-sphere of finite radius \cite{ref:9}.  Various densities of nuclear
material can be explored simply by varying the radius of the
hypersphere.  Since there is only one skyrmion, the calculation is
much more straightforward.  The phase transition may be calculated
analytically.  While the model seems somewhat contrived, since the
nuclear matter density is manifested entirely by the curvature of the
hypersphere, it nevertheless produces similar qualitative and
quantitative (to within a few percent) results as the lattice
calculations at a fraction of the computational cost.  Furthermore, a
study of two skyrmions on a hypersphere \cite{ref:4} suggests that
the simultaneous addition of skymions and the increase in hypersphere
radius, while keeping density fixed, does not significantly affect
the results.  Thus, one may be able to continuously connect the
lattice approach and the hypersphere approach without substantial
change in the observable quantities.

In this paper, we explore kaon condensation by adapting the
bound-state approach to the 3-sphere.  This approach has already been
carried out in \cite{ref:12} using a different version of the Skyrme
model, in which vector mesons are included as fundamental fields.  It
was further reasoned in \cite{ref:12} that the parameter $m_K$, the
mass of the kaon in a vacuum, should ``run'' as a function of the
density.  At the time, this seemed to be the only means in which kaon
condensation could be derived from the Skyrme model.  It was assumed
that unless the energy of the kaon, $\omega$, reaches zero for a
finite value of the 3-sphere radius, kaon condensation would not
occur at all, or at least not for reasonably low densities.  Without
the running kaon mass, $\omega$ reaches zero only at zero radius.
However, with more careful consideration we realize that kaon
condensation sets in at a point where $\omega$ equals the fermi level
of the electrons \cite{ref:2}, which may be much higher than its rest
mass.  This reopened the possibility of using the original version of
the Skyrme model, with no running masses, to calculate the onset of
kaon condensation.  This method has the advantage of requiring fewer
input parameters, thus pinning down a more precise (albeit
model-dependent) transition density.  It is interesting that the
final result lies within the range reported in \cite{ref:13} and lies
fairly close to the values suggested by \cite{ref:12}.

This paper is organized as follows: The bound-state approach is
summarized in Section~\ref{sect:2} and adapted for the hypersphere in
Section~\ref{sect:3}.  The background field is calculated in
Section~\ref{sect:4}.  In Section~\ref{sect:5}, a new discrete
symmetry is found and discussed, which leads the way for the
calculation of the effective kaon mass, $\omega$, as a function of
hypersphere radius in Section~\ref{sect:6}.  Finally, in
Section~\ref{sect:7}, $\omega$ is compared to the electron fermi
level and a prediction for the $K^-$ condensation transition density
is obtained.

\section{The Bound State Approach to the Skyrme Model} \label{sect:2}

A single hyperon may be modeled by the Skyrme model in flat space.
The version used by Callan and Klebanov \cite{ref:6} utilizes a
simple chiral lagrangian with a commutator stabilizing term
\begin{eqnarray}
\sL &=& \frac{f_\pi^2}{16} \tr(\partial_\mu \hc{U} \partial^\mu U)
   + \frac{1}{32e^2} \tr[\partial_\mu U \hc{U},\partial_\nu U \hc{U}]^2
   + \frac{f_\pi^2}{8} \tr \sM(U + \hc{U} - 2) \nonumber \\
 &=& -\frac{f_\pi^2}{16} \tr M_\mu M^\mu
   + \frac{1}{32e^2} \tr[M_\mu,M_\nu]^2
   + \frac{f_\pi^2}{8} \tr \sM(U + \hc{U} - 2), \label{eqn:2.1}
\end{eqnarray}
where $U(\vx,t)\in \SU{3}$. $\sM$ is proportional to the quark
mass matrix and, if we neglect the $u$ and $d$ masses, is given
by
$$\sM = \mtxIIIxIII{0}{0}{\hc{0}}{m_K^2},$$
where, in general, we write $3\times3$ matrices in
the partitioned form
$$\mtxIIIxIII{2\times2}{2\times1}{1\times2}{1\times1}.$$
$M_\mu$ is defined as
$$M_\mu := \partial_\mu U \hc{U}.$$
The standard fit to nucleon and delta masses yields $\fpi=129\,\MeV$
and $e=5.45$.  The empirical value of the kaon mass is $m_K =
495\,\MeV$.

In addition to the local terms, we must also include the Wess-Zumino
term, which is written in non-local form as an integral
over a five-dimensional disk with spacetime as its boundary:
\begin{equation} \label{eqn:2.2}
S_\WZ = -\frac{iN}{240\pi^2} \int_{\disk} d^5 \vx\,
   \epsilon^{\mu\nu\alpha\beta\gamma}
   \tr(M_\mu M_\nu M_\alpha M_\beta M_\gamma),
\end{equation}
where $U(\vx,t)$ is continuously extended to the disk (with our convention,
$\epsilon^{01234}=1$).  $N$ is the number of colors.

In the bound-state approach to the Skyrme model, we use the following
anzatz
\begin{eqnarray} \label{eqn:2.7}
U(\vx,t) &=& \sqrt{U_\pi(\vx,t)}\, U_K(\vx,t) \sqrt{U_\pi(\vx,t)}, \\
U_K(\vx,t) &=&
   \exp i\frac{2\sqrt2}{f_\pi} \mtxIIIxIII{0}{K(\vx,t)}{\hc{K(\vx,t)}}{0},
   \nonumber
\end{eqnarray}
where $U_\pi \in \SU{2}$ and $K$ is a complex spinor.  Substituting this
anzatz into the Skyrme action and expanding to second order in the
kaon fields, we obtain the bound-state lagrangian
\begin{eqnarray}
\sL &=&
-\frac{\fpi^2}{4}\tr(A_\mu^2)
+\frac{1}{2e^2}\tr([A_\mu,A_\nu]^2)
-m_K^2 \hc{K}K
+\hc{D_\mu K}D_\mu K
   \nonumber \\ && \mbox{}
+\frac{6}{e^2 \fpi^2} \hc{D_\mu K}[A_\mu,A_\nu]D_\nu K
-\frac{2}{e^2 \fpi^2} \hc{D_\mu K} D_\nu K \tr(A_\mu A_\nu)
   \nonumber \\ && \mbox{}
+\frac{2}{e^2 \fpi^2} \hc{D_\mu K} D_\mu K \tr(A_\nu^2)
-\frac{2}{e^2 \fpi^2} \hc{K}K \tr([A_\mu,A_\nu]^2)
+\frac12 \hc{K}K \tr(A_\mu^2)
   \nonumber \\ && \mbox{}
+\frac{iN}{3\fpi^2 \pi^2} \epsilon^{\mu\alpha\beta\gamma} 
   \tr(A_\alpha A_\beta A_\gamma) (\hc{D_\mu K}K - \hc{K}D_\mu K),
      \label{eqn:2.3}
\end{eqnarray}
where
\begin{eqnarray*}
D_\mu K &=& \partial_\mu K + V_\mu K, \\
A_\mu &=& \frac12 (
   \hc{\sqrt{U_\pi}\,} \partial_\mu \sqrt{U_\pi} -
   \sqrt{U_\pi} \partial_\mu \hc{\sqrt{U_\pi}\,}
), \\
V_\mu &=& \frac12 (
   \hc{\sqrt{U_\pi}\,} \partial_\mu \sqrt{U_\pi} +
   \sqrt{U_\pi} \partial_\mu \hc{\sqrt{U_\pi}\,}
).
\end{eqnarray*}

The term proportional to $N$ is the result of integrating the
Wess-Zumino term.  The lagrangian itself is quadratic in kaon fields
and describes non-interacting kaons bound to a background \SU{2}
skyrmion.  The background field is realized by the hedgehog anzatz
\begin{equation} \label{eqn:2.4}
U_\pi(\vx,t) = \mtxIIIxIII{e^{iF(r)\,\vtau\cdot\unitr}}{0}{\hc0}{1}
\end{equation}
with the boundary conditions $F(0)=\pi$ and $F(\infty)=0$.  We write
the eigenmodes for $K(\vx,t)$ in terms of its spin and isospin.  In
particular, we will be interested in S-wave and P-wave kaons, for
which the corresponding anz{\"a}tze are
\begin{eqnarray}
   \label{eqn:2.5}
K_S(\vx,t) &=& k_S(r) \chi(t), \\
   \label{eqn:2.6}
K_P(\vx,t) &=& ik_P(r) \,\vtau\cdot\unitr\, \chi(t).
\end{eqnarray}
In each case, $\chi(t)$ are the dynamical variables arranged in a
complex spinor.

The field $F(r)$ is determined classically using the non-kaon part of
the lagrangian.  This baryon field is treated as a background field
with which the kaons interact.  The energy eigenstates form a Fock
space for each kaon mode, where the energy required to create a kaon
in a particular mode is given by its classical eigenfrequency
(i.e.\ substitute $\dot\chi = -i\omega\chi$ and minimize the action
with respect to $\omega$ and $k(r)$).  In flat space, the lowest
energy modes for P-wave and S-wave are given respectively by
$\omega_P=153\,\MeV$ and $\omega_S=368\,\MeV$.

\section{The Skyrme Model on a Hypersphere} \label{sect:3}

In \cite{ref:9}, Manton suggested that a baryon crystal may be
approximated by putting a single Skyrmion on a hypersphere of finite
radius.  This has the effect of allowing the baryon to interact with
itself.  We will continue with this idea by considering the bound
state between a kaon and a baryon on the hypersphere.

A hypersphere of radius $a$ is described by three angular coordinates
$(\rho,\theta,\phi)$ ($0\leq\rho,\theta\leq\pi$ and
$0\leq\phi\leq2\pi$) with the following spatial metric
$$ds^2 = a^2(d\rho^2 + \sin^2\rho (d\theta^2 + \sin^2\theta d\phi^2)).$$
It is often convenient to replace the $\rho$ coordinate with
$x=\cot\rho, -\infty\le x\le \infty$.  This spreads out the
coordinate system around the poles, thereby eliminating coordinate
singularities in the lagrangian.

Topologically, the hypersphere is equivalent to $\reals^3$ with all
the points at infinity mapped to a single point.  However, the metric
is different, with points near infinity ($r\gg a$) much closer
together on the hypersphere than the corresponding points in flat
space.  For $r\ll a$, this difference is negligible.  Since the
baryon size in flat space is on the order $1/e\fpi$, it follows that
for $a \gg 1/e\fpi$, the results on the hypersphere will not be much
different from the results in flat space.  However, for $a$ on the
order of $1/e\fpi$, the skyrmion begins to interact with itself on
the hypersphere, and this can have a significant effect on the
results.

We choose to identify the origin $r=0$ in $\reals^3$ with the south
pole $\rho=\pi$ ($x=-\infty$) and the point at infinity with the
north pole $\rho=0$ ($x=\infty$).  For $r\ll a$, $S^3$ and $\reals^3$
are equivalent, both in topology and metric, with corresponding
points
$$r \approx a(\pi-\rho) \approx -a/x.$$

There are a few technical remarks in adapting the Skyrme model for
the hypersphere.  The profile functions $F$ and $k$ on the
hypersphere will depend on $\rho$ ($x$), which replaces $r$ as the
radial coordinate.  The boundary conditions for $F$ are as follows
$$\begin{tabular}{lcc}
 & at the origin & at infinity \\ \hline
$r$ & $0$ & $\infty$ \\
$\rho$ & $\pi$ & $0$ \\
$x$ & $-\infty$ & $\infty$ \\
Value of $F$ & $\pi$ & $0$
\end{tabular}$$
In choosing to identify the origin with $\rho=\pi$, we reverse the
$\unitrho$ direction compared to the $\unitr$ direction in flat
space.  It is therefore appropriate to reverse the sign of
$\epsilon^{\mu\nu\alpha\beta}$, and hence the Wess-Zumino
term.\footnote{If we choose instead to identify the origin with
$\rho=0$, then $F(0)=\pi$ and $F(\pi)=0$ and the Wess-Zumino term
stays the same.  The results are equivalent.} In flat space, the
matrix $\vtau\cdot\unitr$ depends on the angular coordinates $\theta$
and $\phi$.  On the hypersphere, we identify $\vtau\cdot\unitr$ with
the same function of $\theta$ and $\phi$, except these coordinates
are now the hypersphere coordinates $\theta$ and $\phi$.

Using the anzatz in \refeqn{eqn:2.4} adapted for the hypersphere, we
may compute $A_\mu$ and $V_\mu$.  In terms of the normalized
coordinate basis for the dual tangent space
$$\omega^c = \dc \sqrt{|g_{cc}|}, \qquad c=t,\rho,\theta,\phi,$$
we find $V_\mu$ and $A_\mu$ to given by
\begin{eqnarray*}
V_\mu &=& \frac{1-\cos F}{2r}
   i\vtau\cdot(\unitphi\,\omega^\theta_\mu - \unittheta\,\omega^\phi_\mu), \\
A_\mu &=& \frac{1}{2a}F'(\rho) i\vtau\cdot\unitr\, \omega^\rho_\mu +
   \frac{\sin F}{2r} i\vtau\cdot
   (\unittheta\,\omega^\theta_\mu + \unitphi\,\omega^\phi_\mu).
\end{eqnarray*}
where $r=r_{\rm phys}=a\sin\rho$.  Using this result and
\refeqn{eqn:2.6}, we obtain the (spatially integrated) lagrangian for
a P-wave kaon
\begin{eqnarray}
L &=&
\Flagprefactor\Bigg\{
\FlagtermI
\FlagtermII
\FlagtermIII
  \nonumber \\ && \mbox{}
\FlagtermIIII
\FlagtermV
\FlagtermVI
\Bigg\}
  \nonumber \\ && \mbox{}
+ \klagprefactor\Bigg\{
\klagtermI
\klagtermII
\klagtermIII
  \nonumber \\ && \mbox{}
\klagtermIIII
  \nonumber \\ && \mbox{}
\klagtermV
\klagtermVI
\klagtermVII
\klagtermVIII
\klagtermVIIII
  \nonumber \\ && \mbox{}
\klagtermX
\klagtermXI
\klagtermXII
\klagtermXIII
  \nonumber \\ && \mbox{}
\klagtermXIIII
\klagtermXV
\klagtermXVI
\klagtermXVII
  \nonumber \\ && \mbox{}
\klagtermXVIII
\klagtermXVIIII
\klagtermXX
  \nonumber \\ && \mbox{}
\klagtermXXI
\Bigg\},
   \label{eqn:3.1}
\end{eqnarray}
where $\alpha=ae\fpi$ and we use the substitution $x=\cot\rho$
mentioned earlier.  One could use \refeqn{eqn:2.5} instead of
\refeqn{eqn:2.6} to obtain the lagrangian for S-wave kaons.  However,
as we will see later, it is possible to obtain this lagrangian from
the P-wave lagrangian by substituting $F(x)\mapsto F(x)+\pi$.  The
ramifications of this result will be explored later.

The remainder of the paper will be spent solving the
stationary conditions for $F(x)$ and $k(x)$, and determining
$\omega$ as a function of $a$.

\section{Finding the Background SU(2) Baryon Field} \label{sect:4}

The \SU{2} baryon field is determined by the profile function $F(x)$,
which is obtained by solving the stationary conditions for $F$ using
the non-kaon part of the lagrangian in \refeqn{eqn:3.1}.  Much of
this work has been carried out by Manton \cite{ref:9} using different
methods.

The differential equation for $F(x)$ is given by
\begin{eqnarray}
0 &=&
\FeqntermI
\FeqntermII
\FeqntermIII
\FeqntermIIII
  \nonumber \\ && \mbox{}
\FeqntermV
\FeqntermVI
  \nonumber \\ && \mbox{}
\FeqntermVII.
   \label{eqn:4.1}
\end{eqnarray}
The boundary conditions for $F$ are given by
$$F(-\infty) = \pi, \qquad F(\infty) = 0.$$

The identity profile function $F(x) = \arccot(x)$ ($F(\rho)=\rho$) is
a solution of \refeqn{eqn:4.1} for every $\alpha$.  It corresponds to
a baryon uniformly distributed on the hypersphere.  For $\alpha$ less
than the critical value $\alpha_{\rm cr} = 2\sqrt2$ it is the only
solution and it is stable.

As $\alpha$ crosses the critical value, the identity solution becomes
unstable in favor of an asymmetric solution, where the baryon is
located at one of the poles.  There are actually two equivalent
solutions which correspond to each other under the transformation
$$F(x) \mapsto \pi-F(-x),$$
which leaves the differential equation invariant.  We always choose
the solution with the baryon localized around $x=-\infty$ (where
$F=\pi$).\footnote{Localizing the baryon around the $F=\pi$ pole, as
opposed to the $F=0$ pole, results in a slightly lower energy for the
physically relevant case $m_\pi \neq 0$.}

To find the localized solution, we numerically solve the differential
equation using the initial conditions provided by the asymptotic
solution
$$F(x) \approx \pi + A/x + \ldots \qquad x \ll 0$$
and use trial and error with the value of $A$ until we find a solution
which approaches zero as $x\rightarrow \infty$.
For $a\gg 1/ef_\pi$ (i.e.\ $\alpha \gg 1$), we expect $F(x)$ to approach
the solution corresponding to flat space.  In particular, $F$ takes its
middle value ($\pi/2$) at around $r=a\sin\rho \sim 1/ef_\pi$ which
corresponds to $x\sim -\alpha$.  Thus $A\sim\alpha$ for large $\alpha$.

\section{A Discrete Symmetry} \label{sect:5}

In section~\ref{sect:4} we found that the non-kaon part of the
lagrangian is symmetric under the transformation
\begin{equation} \label{eqn:5.1}
F(x) \mapsto \pi-F(-x).
\end{equation}
This transformation maps the baryon from one pole to the other.  A
simple coordinate transformation corresponding to point inversion,
which also maps the baryon from one pole to another, is given by
\begin{eqnarray}
\phi &\mapsto& \phi + \pi, \nonumber \\
\theta &\mapsto& \pi-\theta, \nonumber \\
x &\mapsto& -x, \nonumber \\
   \label{eqn:5.2}
F(x) &\mapsto& -F(-x),
\end{eqnarray}
and leaves the original Skyrme lagrangian invariant.\footnote{The
boundary conditions on $F(x)$ are different.} \refeqn{eqn:5.1}
differs from the coordinate inversion of \refeqn{eqn:5.2} by the
transformation
\begin{equation} \label{eqn:5.3}
F(x) \mapsto F(x)+\pi.
\end{equation}

If one examines closely, one realizes that the kaon part of the
P-wave lagrangian in \refeqn{eqn:3.1} is not symmetric under
\refeqn{eqn:5.1} or \refeqn{eqn:5.3}, regardless of the substitution
for $k(x)$, although it is symmetric under \refeqn{eqn:5.2}.
However, \refeqn{eqn:5.3} (and hence \refeqn{eqn:5.1}) corresponds to
a valid symmetry of the bound-state lagrangian.  It is no accident
that the non-kaon part is symmetric under \refeqn{eqn:5.1}.  To
determine what this symmetry transformation does to the kaon field,
we need to examine the effect of \refeqn{eqn:5.3} on the Skyrme
lagrangian.

The transformation in \refeqn{eqn:5.3} corresponds to
$$U_\pi \mapsto \mtxIIIxIII{-\id}{0}{\hc0}{1} U_\pi.$$
The bound-state approach requires the value of $\sqrt{U_\pi}$.
Technically, the square root is ambiguous because $U_\pi$ is not
positive-definite hermetian.  We define the square root to be
$$\sqrt{U_\pi} =
   \mtxIIIxIII{e^{i\frac{F(x)}{2}\,\vtau\cdot\unitr}}{0}{\hc0}{1}.$$
Hence \refeqn{eqn:5.3} corresponds to
$$\sqrt{U_\pi}
   \mapsto \mtxIIIxIII{i\vtau\cdot\unitr}{0}{\hc0}{1} \sqrt{U_\pi}
   = \sqrt{U_\pi} \mtxIIIxIII{i\vtau\cdot\unitr}{0}{\hc0}{1}.$$
If we arrange for the kaon field to transform so that
\begin{equation} \label{eqn:5.4}
U_K \mapsto \hc{W} U_K W, \qquad
   W := \mtxIIIxIII{i\vtau\cdot\unitr}{0}{\hc0}{1},
\end{equation}
then $U(\vx,t)$ itself will transform as
\begin{equation} \label{eqn:5.5}
U \mapsto \mtxIIIxIII{-\id}{0}{\hc0}{1} U.
\end{equation}
The Skyrme lagrangian \refeqn{eqn:2.1} is invariant under
\refeqn{eqn:5.5}.  The transformation in \refeqn{eqn:5.4} corresponds
to
\begin{equation} \label{eqn:5.6}
K \mapsto -i\vtau\cdot\unitr\,K.
\end{equation}
If we compare the anz{\"a}tze in \refeqn{eqn:2.5}
and~\refeqn{eqn:2.6} we find that, aside from a trivial change in the
sign of $k(x)$, \refeqn{eqn:5.6} corresponds to the exchange in the
roles of S-wave and P-wave.  The crucial conclusion from the
discussion above is that the simultaneous transformations
\refeqn{eqn:5.1} and~\refeqn{eqn:5.6} leave the action
invariant.\footnote{This symmetry is broken by introducing a pion
mass, which we neglect in this paper.} This simultaneous
transformation does the following:
\begin{itemize}
\item It moves the baryon from one pole to the other.

\item It does {\em not\/} affect the boundary conditions of $F(x)$.
Therefore, the value of $F$ at the pole which the baryon is nearest
to changes from $\pi$ to $0$ or vice-versa.

\item It transforms the kaon from S-wave to P-wave, but does not
otherwise move the kaon ($k(x)$ does not change, except for an
irrelevant change in sign).
\end{itemize}

For $\alpha>\alpha_{\rm cr}$ the solution for $F(x)$ does not respect
\refeqn{eqn:5.1}.  In this phase, the kaon field is not expected to
respect~\refeqn{eqn:5.6} either.  Thus, there is no symmetry between
the S-wave and P-wave kaons.  However, for $\alpha<\alpha_{\rm cr}$,
the solution $F(x)=\arccot(x)$ ($F(\rho)=\rho$) does
respect~\refeqn{eqn:5.1}.  In this phase, {\em there is no
distinction between S-wave and P-wave kaons}.  In the next section,
we see the value of $\omega$ for S-wave and P-wave approach each
other as the critical value of $\alpha$ is approached.  For
$\alpha<\alpha_{\rm cr}$, there is only one value of $\omega$.

\section{Solving for the Kaon Field} \label{sect:6}

The next goal is to solve the kaon part of the lagrangian in
\refeqn{eqn:3.1} for $\omega$ and the kaon field as a function of
$\alpha$.  In the next section, we will compare the electron fermi
level to $\omega$ and determine the value of $\alpha$ where kaon
condensation sets in.

In the non-uniform phase $\alpha>2\sqrt2$, we substitute the
numerical solution to $F(x)$ into the Euler-Lagrange equations
corresponding to the lagrangian and numerically solve for $k(x)$ and
$\omega$.  We do this separately for P-wave and S-wave.

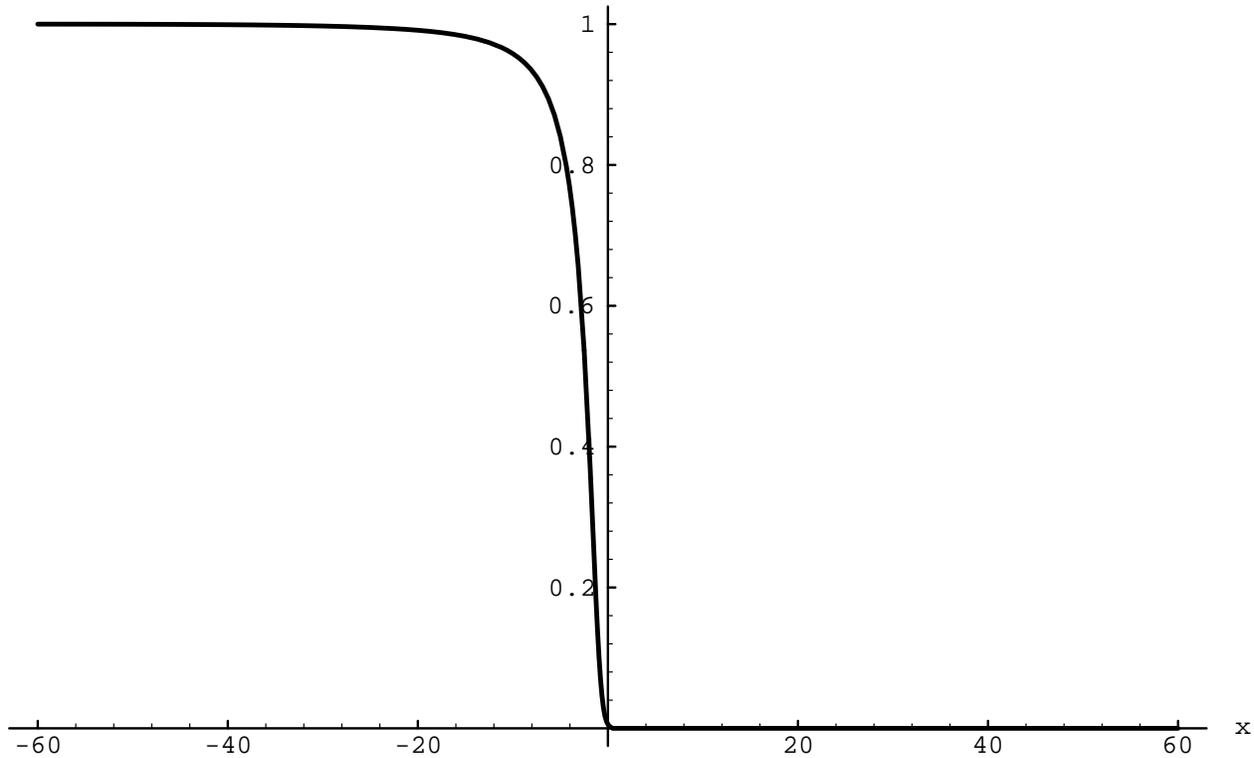
\begin{figure}[e]
\caption{P-wave kaon wavefunction for $\alpha=5$} \label{fig:6.2}
\end{figure}

For P-wave kaons, we use the lagrangian given in \refeqn{eqn:3.1}.  The
initial conditions are provided by the asymptotic solution
$$k(x) \approx k_0 + \ldots \qquad x\ll0.$$
Since the differential equation is linear in $k(x)$, we may fix the
value of $k_0=1$.  The value of $\omega$ is adjusted until we find a
solution which approaches zero as $x\rightarrow\infty$.  A typical
(unnormalized) solution for $k(x)$ is given in Fig.~\ref{fig:6.2}.

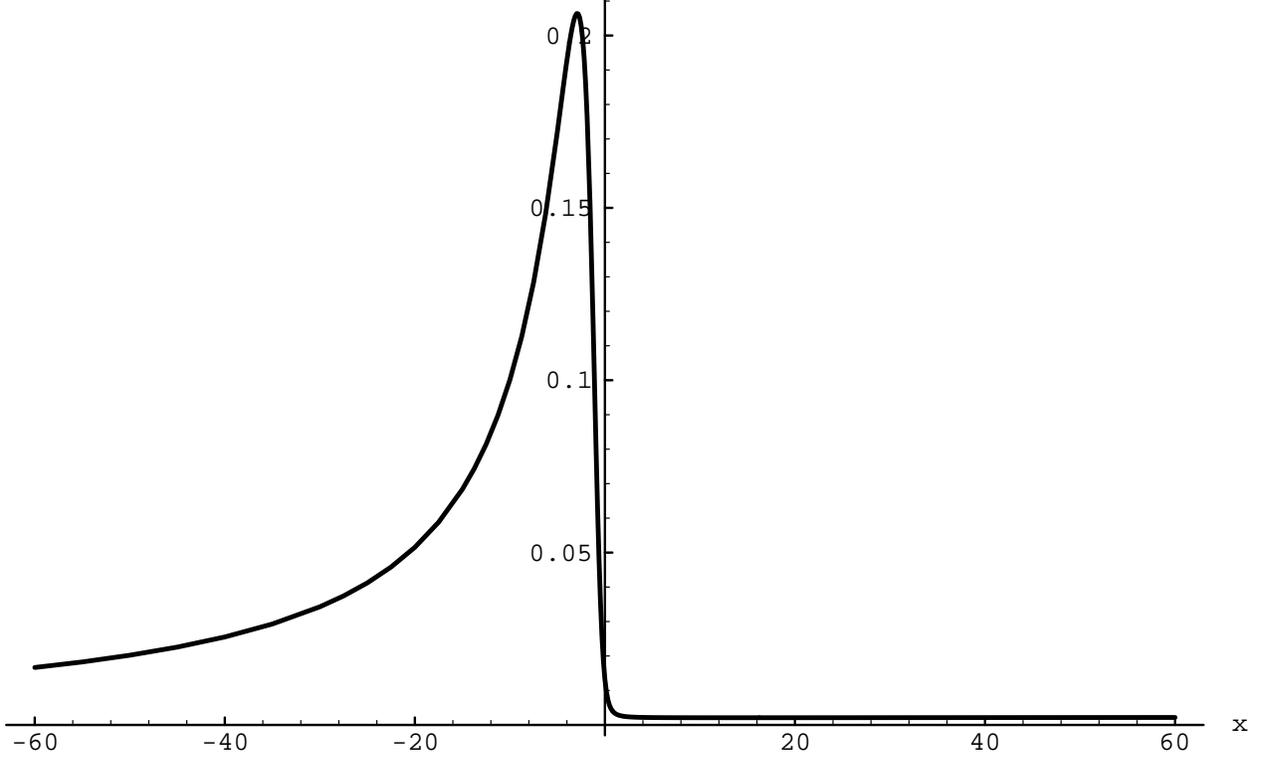
\begin{figure}[e]
\vspace{1in}
\caption{S-wave kaon wavefunction for $\alpha=5$} \label{fig:6.3}
\end{figure}

For S-wave kaons, we use the S-wave lagrangian obtained from
\refeqn{eqn:3.1} by the substitution $F(x)\mapsto F(x)+\pi$.  This
transformation differs from~\refeqn{eqn:5.1} by coordinate inversion,
and so the kaon switches poles instead of the baryon.  The initial
conditions are provided by the asymptotic solution
$$k(x) \approx -A/x + \ldots \qquad x\ll 0.$$
We fix $A=1$ and adjust the value of $\omega$ until we find a
solution which approaches a constant as $x\rightarrow\infty$.  A
typical (unnormalized) solution for $k(x)$ is given in
Fig.~\ref{fig:6.3}.

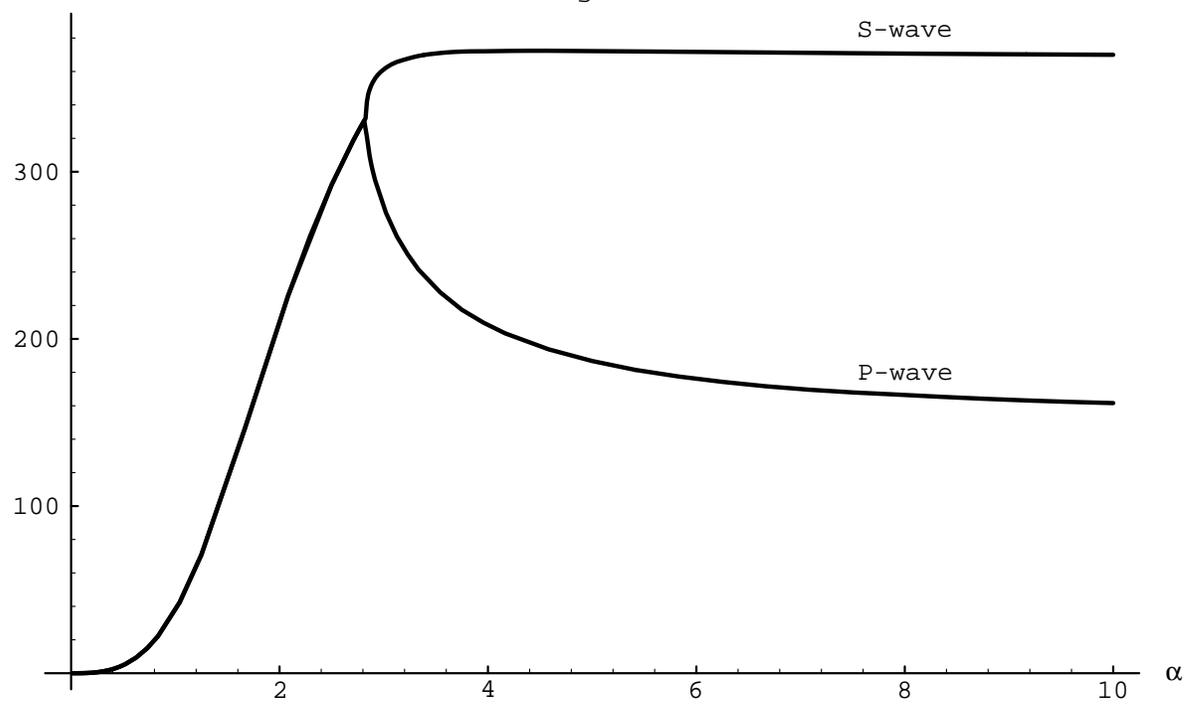
\begin{figure}[e]
\vspace{1in}
\caption{kaon energy (MeV) vs.\ $\alpha$} \label{fig:6.1}
\end{figure}

The solution for $\omega$ as a function of $\alpha$ for both S-wave
and P-wave is given in Fig.~\ref{fig:6.1}.

In the uniform phase $\alpha < 2\sqrt2$, we substitute the analytic
solution $F(x)=\arccot x$ ($F(\rho)=\rho$) into \refeqn{eqn:3.1}.  As
explained in the previous section, there is no distinction between
S-wave and P-wave kaons in the uniform phase.  If we substitute the
same analytic form for $F(x)$ into the lagrangian appropriate for
S-wave, we obtain the same differential equation, except for the
coordinate inversion $x\mapsto -x$.

The resulting differential equation is
\begin{equation} \label{eqn:6.1}
k''(x) + \frac{k(x)}{(1+x^2)^2}(\frac14 - x\sqrt{1+x^2} - x^2 + \lambda) = 0,
\end{equation}
where
\begin{equation} \label{eqn:6.2}
\lambda =
  \frac{3+\alpha^2}{2+\alpha^2} \parens{\frac{\omega\alpha}{e\fpi}}^2
  + \frac{e^2 N}{\pi^2(2+\alpha^2)} \frac{\omega\alpha}{e\fpi}
  - \frac{1}{2+\alpha^2} \parens{\frac{m_K}{e\fpi}}^2 \alpha^4.
\end{equation}

The dependence of $\omega$ on $\alpha$ is entirely contained in the
expression for $\lambda$.  In the introduction, we claimed that
$\omega\rightarrow 0$ as $\alpha\rightarrow 0$.  From
\refeqn{eqn:6.2}, it is clear that this is only possible if
there exists a solution to \refeqn{eqn:6.1} for $\lambda=0$.  This is
not a trivial prediction and there is no reason to expect it.
However, if we rewrite \refeqn{eqn:6.1} in terms of $\rho$ instead of
$x$, we do in fact find a solution for $\lambda=0$.  It
is\footnote{This solution was also found in a different version of
the Skyrme model \cite{ref:12}.}
\begin{equation} \label{eqn:6.6}
k(\rho) = \sin \frac12 \rho.
\end{equation}
It is this miracle which ultimately is responsible for the fact that
the Skyrme model yields kaon condensation at reasonable densities, if
at all.  Otherwise, $\omega \sim \lambda/\alpha$, which is the same
$\alpha$-dependence as the electron fermi energy.  It would not be
clear for what values of $\alpha$, if any, that $\omega \le
\epsilon^e_F$.

Solving \refeqn{eqn:6.2} with $\lambda=0$ for $\omega$ as a function
of $\alpha$, we find\footnote{The other solution to the quadratic
equation (corresponding to a minus sign in front of the square root
sign) corresponds to $\omega<0$ which represents the energy of a
$K^+$ in nuclear matter.  As expected, $|\omega|$ diverges as
$N/\alpha$ as $\alpha\rightarrow0$ due to the Wess-Zumino term.}
\begin{equation} \label{eqn:6.3}
\frac{\omega\alpha}{e\fpi} = \frac{
 \sqrt{
   (\frac{e^2 N}{2\pi^2})^2 + (\frac{m_K}{e\fpi})^2 \alpha^4(3+\alpha^2)
 } - \frac{e^2 N}{2\pi^2}
}{3+\alpha^2}.
\end{equation}
After plugging in numbers, we get
\begin{equation} \label{eqn:6.4}
\omega = (3174\,\MeV)
   \frac{\sqrt{1+.02433 \alpha^4(3+\alpha^2)} - 1}{\alpha(3+\alpha^2)}.
\end{equation}
At the critical point, we find $\omega = 332\,\MeV$, which is consistent
with the numerical solutions for $\alpha>2\sqrt2$.  For
$\alpha\rightarrow0$, we find
\begin{equation} \label{eqn:6.5}
\omega \approx (38.6\,\MeV)\alpha^3
    (1+\order \frac{\alpha^4(3+\alpha^2)}{40}).
\end{equation}

It should be noted that $k(\rho)$ given in~\refeqn{eqn:6.6} satisfies
$$k(\rho) = \sin \frac12 F(\rho).$$
This is the form for $k(\rho)$ which corresponds to a small rigid
rotation into the strange directions \cite{ref:6}.  Thus the rigid
rotator approximation to the bound-state approach given
in~\cite{ref:5} becomes exact in the uniform phase.  Indeed, the
formula~\refeqn{eqn:6.3} for $\omega$ can be expressed in the form
obtained in~\cite{ref:5} using the rigid rotator approximation:
$$\omega = \frac{N}{8\Phi}\parens{\sqrt{1+(\frac{m_K}{M_0})^2} - 1}, \qquad
   M_0 = \frac{N}{4\sqrt{\Gamma\Phi}}.$$
The appropriate values for $\Phi$ and $\Gamma$ are
$$\Phi=\frac{\pi^2 \alpha(3+\alpha^2)}{4e^3 \fpi}, \qquad
   \Gamma=\frac{\pi^2 \alpha^3}{e^3 \fpi}.$$
In light of the fact that $k(\rho)$ as given in~\refeqn{eqn:6.6} is
independent of $m_K$, it is no surprise that the form of $k(\rho)$
matches the rigid rotator form.  For $m_K=0$, we expect the \SU{3}
flavor symmetry to be restored.

The only remaining question is whether or not $\lambda=0$ actually
yields the lowest value of $\omega$ for {\em all\/} $\alpha$ below
the critical value.  We don't expect $\lambda\neq 0$ to yield lower
energy solutions for any $\alpha < 2\sqrt2$ because
\begin{itemize}
\item $\omega(\alpha_{\rm cr})$ is predicted correctly by
$\lambda=0$.  Also, the $\lambda=0$ solution for $k(x)$ matches the
numerical solutions we obtain for $k(x)$ for $\alpha>2\sqrt2$ for
both S-wave (inverted) and P-wave.
\item $k(\rho)$ given by the $\lambda=0$ solution contains no nodes,
which is what we expect for the ground state solution.
\item $k(\rho)$ given by the $\lambda=0$ solution matches the rigid
rotator solution, a situation we expect for $m_K=0$.
\item $\lambda=0$ must yield the lowest $\omega$ for $\alpha$
sufficiently close to zero.  All other values of $\lambda$ result in
a divergent $\omega$ as $\alpha\rightarrow0$.
\item A sudden jump from $\lambda=0$ to a different value of
$\lambda$ would involve a discontinuous jump in $\omega$ and in
$k(\rho)$, neither of which is expected.
\end{itemize}

\section{Determining the Onset of Kaon Condensation} \label{sect:7}

To study the onset of kaon condensation, we need to understand what
is going on in the neutron star.  In our model, the neutron star
consists of a Skyrme lattice of protons and neutrons, as well as
enough electrons to balance the charge of the protons.  For
simplicity, we shall treat these particles as free particles.
We assume that the star is in quasi-equilibrium.  In particular, the
following reaction
$$n \mapsto p^+ + e^- + \bnu$$
is in equilibrium.  This together with conservation of baryon number
and charge allows us to solve three equations
\begin{eqnarray*}
\epsilon^n_F &=& \epsilon^p_F + \epsilon^e_F, \\
\rho_B &=& \rho_p + \rho_n, \\
\rho_p &=& \rho_e,
\end{eqnarray*}
for the three variables $\rho_n$, $\rho_p$, $\rho_e$, in terms of
$\alpha$.  The baryon number density is given by the inverse volume
of the hypersphere
$$\rho_B = (2\pi^2 a^3)^{-1}.$$
The fermi energy $\epsilon^f_F$ for a spin $1/2$ fermion is given in
terms of its number density by
\begin{eqnarray}
\epsilon^f_F &=& \sqrt{m_f^2 + k_F^2} \nonumber \\
 &=& \sqrt{m_f^2 + (3\pi^2 \rho_f)^{2/3}}. \label{eqn:7.1}
\end{eqnarray}

This will allow us to calculate $\epsilon^e_F$ as a function of
$\alpha$.  Before doing so, we wish to discuss kaon condensation.  In
free space, the conversion of electrons to kaons via the reaction
$$e^- \mapsto K^- + \nu$$
is unthinkable, because the mass of the kaon is much larger than that
of the electron.  However, from the previous section we see that the
effective mass of the kaon, $\omega$, approaches zero for large
baryon densities.  Furthermore, for large baryon densities, the fermi
level of the electrons (as well as the protons and neutrons)
increases.  It follows that there exists a value of $\alpha$ for
which $\epsilon^e_F = \omega$.  Beyond that density, it becomes
favorable for electrons to become kaons.  Since kaons are bosons,
there is no fermi level for the kaons.  Every kaon requires the same
energy $\omega$ to produce.  As explained in \cite{ref:15}, kaon
condensation on the order of a few times normal nuclear matter
density may have an impact on the formation of low mass black holes,
by reducing the resistance of the neutron star to collapse.  We now
compute the density of kaon condensation as predicted by our model.

The formula in \refeqn{eqn:7.1} for the fermi energy is fully
relativistic.  We shall make the approximation that the fermi-level
electrons are ultra-relativistic and the fermi-level nucleons are
non-relativistic.  Thus the fermi energies are given by
$$\epsilon^{(p,n)}_F = m_N + \frac{(3\pi^2 \rho_{p,n})^{2/3}}{2m_N},
  \qquad \epsilon^e_F = (3\pi^2 \rho_e)^{1/3}.$$
Conservation of charge and baryon number allows us to express
$\rho_n$, $\rho_p$ and $\rho_e$ in terms of one parameter,
$\gamma=\rho_p/\rho_B$.
$$\rho_p = \rho_e = \gamma \rho_B, \qquad \rho_n = (1-\gamma)\rho_B.$$
In terms of this parameter, the fermi levels may be reexpressed
\begin{eqnarray*}
\epsilon^n_F &=& (3\pi^2\rho_B)^{1/3} \beta (1-\gamma)^{2/3} + m_N, \\
\epsilon^p_F &=& (3\pi^2\rho_B)^{1/3} \beta \gamma^{2/3} + m_N, \\
\epsilon^e_F &=& (3\pi^2\rho_B)^{1/3} \gamma^{1/3}, \\
\beta &:=& \frac{(3\pi^2\rho_B)^{1/3}}{2m_N}.
\end{eqnarray*}
Plugging in numbers yields
\begin{eqnarray*}
(3\pi^2\rho_B)^{1/3} &=& (805/\alpha)\,\MeV, \\
\beta &=& 0.429/\alpha.
\end{eqnarray*}
The equilibrium equation then becomes
\begin{equation} \label{eqn:7.2}
\beta((1-\gamma)^{2/3} - \gamma^{2/3}) = \gamma^{1/3},
\end{equation}
which we can solve for $\gamma$ in terms of $\alpha$.

To get an order of magnitude estimate, we calculate the fermi
energies at $\alpha=2\sqrt2$.  This corresponds to $\beta=0.152$.
Solving~\refeqn{eqn:7.2} iteratively for $\beta$ small, we find
$$\gamma = \beta^3 - 3\beta^5 + \ldots \approx 3.27\times10^{-3}$$
from which we compute the fermi levels
\begin{eqnarray*}
\epsilon^e_F(\alpha_{\rm cr}) &=& 42.2\,\MeV, \\
\epsilon^p_F(\alpha_{\rm cr})-m_N &=& 0.95\,\MeV, \\
\epsilon^n_F(\alpha_{\rm cr})-m_N &=& 43.2\,\MeV.
\end{eqnarray*}
The relativistic approximations made earlier are justified by these
results.  Note that $\omega(\alpha_{\rm cr}) = 332\,\MeV$ is much
larger than $\epsilon^e_F$ at the phase transition, so that kaon
condensation must take place for $\alpha<2\sqrt2$.  However, $\omega$
decreases quickly from its critical value and $\epsilon^e_F$
continues to rise.\footnote{In the uniform phase, the nucleons lose
their identity, so the calculation of the fermi level might seem
suspicious in this regime.  Nevertheless, it is reasonable to expect
the fermi level to continue to rise and for~\refeqn{eqn:7.1} to be a
reasonable approximation.} For $\alpha=1$, $\omega\approx 40\,\MeV$,
which is definitely lower than the fermi level at $\alpha=1$.
Therefore, kaon condensation occurs somewhere in the regime
$1<\alpha<2\sqrt2$.

To find the transition exactly, we note that~\refeqn{eqn:7.2} can be
{\em explicitly\/} solved for $\alpha$ in terms of $\gamma$:
$$\alpha = (.429)\frac{(1-\gamma)^{2/3} - \gamma^{2/3}}{\gamma^{1/3}}.$$
Hence both sides of the equation $\epsilon^e_F = \omega$ can be
written explicitly in terms of $\gamma$.  Numerical solution yields
$\gamma=0.016$.\footnote{This means protons make up 1.6 percent of
the nucleons in the neutron star.  Most of the protons have already
recombined with electrons to form neutrons.} The following values are
found at the onset of kaon condensation
\begin{eqnarray*}
\gamma &=& 0.016, \\
\alpha &=& 1.58, \\
a &=& 0.44\,\fm, \\
\omega = \epsilon^e_F &=& 129\,\MeV, \\
\epsilon^p_F-m_N &=& 8.8\,\MeV, \\
\epsilon^n_F-m_N &=& 138\,\MeV, \\
\rho_B &=& 0.595\,\fm^{-3}.
\end{eqnarray*}
The density for kaon condensation is on the order of a few (3--4)
times the density of normal nuclear matter
$\rho_{\rm nucl}=0.16\,\fm^{-3}$.

\section{Conclusions} \label{sect:8}

We have predicted the onset of kaon condensation to be
$\rho_B=0.595\,\fm^{-3}=3.7\rho_{\rm nucl}$.  We offer a brief
comparison to the literature in Table~\ref{tab:8.1}.
\begin{table}
\caption{Values of the baryon transition density $\rho_B$} \label{tab:8.1}
\begin{center}
\begin{tabular}{lcc}
 & Transition density & Relative density \\
Source & $\rho_B$ & $\rho_B/\rho_{\rm nucl}$ \\ \hline
Bound-state approach & $0.595\,\fm^{-3}\vphantom{\sqrt{x^2}}$ & $3.7$ \\
on a hypersphere &&\\
Mean field approach & $0.3$--$0.6\,\fm^{-3}$ & $2$--$4$ \\
with chiral lagrangian \cite{ref:13} &&\\
Vector-meson Skyrme & $0.15$--$0.23\,\fm^{-3}$ & $0.9$--$1.4$ \\
model on hypersphere \cite{ref:12} &&
\end{tabular}
\end{center}
\end{table}
Beyond the critical density, an increasing fraction of neutrons is
converted to protons and $K^-$, which softens the star's resistance to
compression.  As explained in \cite{ref:15}, a critical density on
the order of a few times the nuclear density is sufficient to reduce
the upper mass limit beyond which the collapse of a star to a black
hole is inevitable.  Our results fall within that range.

The key ingredient which allows the bound-state approach to predict a
reasonable transition density is using the electron fermi level.
Without the fermi level, kaon condensation would not set in until
infinite density.

Now that it has been shown that the Skyrme model can produce
reasonable results for $K^-$ condensation, several refinements of the
calculation are possible.
Since the rigid rotator approximation to the bound-state approach
becomes exact for $\alpha<2\sqrt2$, we may calculate the complete
$1/N$ correction to the $K^-$ energy using the formulae generated in
\cite{ref:17}.  A more ambitious project is to replace the
hypersphere with a real lattice.  In addition, a more sophisticated
model of the neutron star may be used to determine the electron fermi
energy more precisely.  None of these refinements is expected to
drastically change the results.

\section*{Acknowledgements}

I would like to thank I. R. Klebanov, my thesis advisor, for
recommending this problem to me and for sharing his ideas.
This work was supported in part by DOE grant DE-FG02-91ER40671,
the NSF Presidential Young Investigator Award PHY-9157482, and
James S. McDonnell Foundation grant No. 91-48.

\section*{Figures}

\begin{description}
\item[Fig.\ \ref{fig:6.2}]
P-wave kaon wavefunction (unnormalized) for $\alpha=5$.

\item[Fig.\ \ref{fig:6.3}]
S-wave kaon wavefunction (unnormalized) for $\alpha=5$.

\item[Fig.\ \ref{fig:6.1}]
Kaon energy (MeV) vs.\ $\alpha$.
\end{description}

\end{document}